# Evaluating the Effect of Credit Collection Policy on Portfolio Quality of Micro-Finance Bank

## Esther Yusuf Enoch, Abubakar Mahmud Digil and Usman Abubakar Arabo

Corresponding Author: Esther Yusuf Enoch
*Department of Banking and Finance,*
*Federal Polytechnic Mubi, Nigeria*

**ABSTRACT:** *This study evaluates the effect of collection policy on portfolio quality of microfinance banks in Adamawa State, Nigeria. Real data were collected from 51 credit officers, then a multi-stage sampling method was used to select a sample of 21 respondents from the population (i.e., 51 credit officers). In addition, we used regression analysis and descriptive statistics to analyze the data collected and to also test our proposed hypothesis. Based on the evaluation performed, the results showed that collection policy has a higher effect on portfolio quality. Hence, the study showed that microfinance banks should adhere to strict or stiff debt collection policy as strictness in collection policy help the banks to recover their loans, thereby improving the portfolio quality of the bank.*

---------------------------------------------------------------------------------------------------------------------------------

---------------------------------------------------------------------------------------------------------------------------------

## I. INTRODUCTION

Credit management is an important activity that cannot be overlooked by any financial institution irrespective of its business nature. It is the process of controlling and collecting payments from customers for the service rendered to them. Credit management can be seen as the methods and strategies employed by a firm to maintain an optimal level of credit (Myers and Brealey, 2003). Good credit management will help reduce the amount of capital tied up with debtors and minimize your exposure to bad debts.

Credit management refers to the process of lending starting from inquiring potential borrowers up to recovering the amount granted. In the banking industry, credit management involves accepting applications, client appraisal, loan approval, monitoring, and recovering of these loans Credit management is vital for any entity dealing with the grating of loans since it is impossible to have a zero credit or default risk. Nelson (2002) opined that credit management deals with the way an entity manages its credit sales effectively and efficiently. A weak credit management system automatically leads to business failure. Charles (1999). Stressed the importance of credit management as follows; Credit management process deserves special emphasis because proper credit management greatly influences the success or failure of financial institutions. Therefore, the need to employ proper credit policies and procedures that enhance the performance of credit management and protect the banking industry from failure. According to Hettihewa (1997). Credit Management is important as granting credit is considered to be equivalent to investing in a customer. However, payment of debts should not be delayed because bad debts are a cost to a firm as credit management is concerned with managing debtors and financing debts.

### 1.2 Related Work

In this section, we provide the related works on collection policy, portfolio quality of microfinance banks, financial performance, microfinance banks, and credit management.

1.2.1 Collection Policy and Portfolio Quality of Microfinance Banks

An effective collection policy strategy begins with a spelled out credit policy and credit management tools for enforcing the policy. Success comes from the overall performance of the whole credit value chain. Collection policies within financial institutions can make the difference between a good and excellent performance for the bank, through making use of opportunities to make the collection processes strategically effective, operationally efficient, and customer-oriented (Benveniste, 2003). A debt collection policy can be defined as a legitimate and necessary business activity where creditors and collectors are able to take reasonable steps to secure payment from customers who are legally bound to pay or repay the money they owe (Kitua, 2002). Once a loan or credit agreement has been finalized and paid out to a customer, the next phase of the credit provider's task will start. The credit agreement has to be actively managed over its life cycle up to the payment dates when customers are expected to fully pay their debt. However, as a result of various reasons the payment agreement may not usually be honored as anticipated, this could lead to late repayment (Benveniste, 2003). A sound collection policy is paramount because some clients are slow payers while some do not pay at all. Collection efforts should therefore be aimed at accelerating collections from slow and non-payers in order to reduce bad debt and for the bank to have enough funds to meet up with its customer's demands.

Portfolio quality deals with how well or how best a microfinance institution can protect this portfolio (loans) against all forms of risk. Portfolio Quality entails the total fund available for the MFIs to use as loans to satisfy a client. The availability of funds by MFIs depends largely on the collection policy; if the collection policy is effective and properly applied the portfolio quality automatically increases likewise the customers' satisfaction, on the other hand, overdue payment will have a negative impact on the credit provider's financial performance, since cash flow targets cannot be met and collection cost increases. Risk is also indicated by overdue payments, taking into account that overdue payment may become payment that cannot be collected and inevitably resulting in losses and bad debt. As a result of all these negative consequences, the collection of overdue accounts is extremely important to any credit provider (Harvey, 2005) Weak collection policy could mean poor asset quality while a sound collection policy ensures high asset quality. The management's ability to recover loans disbursed to its client's results to fat loan portfolio, with effective measures put in place by MFIs to ensure the full recovery of loans, banks will record-high profit.

The loan portfolio is not only considered as the largest asset or pre-dominant source to generate revenue but one of the biggest risk sources for the financial institution's soundness and safety as well (Richard, Chijioriga, Kaijage, Perterson and Bohman 2008). Hence, credit risk management is considered to be one of the road maps for soundness and safety of the sector through prudent actions as well as monitoring performance

### 1.2.2 Financial Performance

Financial performance involves measuring the results of a firm's policies and operations in monetary terms. It is used to measure a firm's overall financial health over a given period of time.

The ability to survive, grow, operate efficiently and profitably is known as financial performance (Turyahebya, 2013). The resources use to achieve the firm objectives can be used in measuring the financial performance of that firm. Therefore, a firm's low performance can be due to poor performing assets.

The financial performance consists of how efficiently a bank uses its assets from its primary mode of business and generates revenues. Banks, as the critical part of the financial system, play a vital role in contributing to a country's economic development. The financial performance of banks can be measured using ratios such as profitability, efficiency, and portfolio ratio.

### 1.2.3 Micro Finance

Micro Finance refers to the provision of financial services, primarily savings and credit to the poor and low-income households that do not have access to commercial bank service. CGAP (2012) defined "Micro Finance" as the provision of formal financial services to poor and low-income people, as well as others systematically not benefiting from the financial system. Microfinance does not only, provide a range of credit products for consumption, smoothing for business purposes, to fund social obligations, for emergencies, etc, only but also savings, money transfers, and insurance.

Micro Finance is an attempt to improve access to small deposits and small loans for poor households neglected by commercial banks. Micro Finance is one of the ways of building the capabilities of the poor who are largely ignored by commercial banks and other lending institutions and graduating them to sustainable self-employment activities by providing them financial services like credit, savings, and insurance (Annad and Kanwal, 2011). The reason for this neglect is many. Often, such credits are just not profitable enough for banks, because of the economic scale. By focusing on a small amount, and easing collateral requirements, microfinance institutions are better equipped to target poor individuals or groups who need resources to finance small-scale investment.

These credits can be sufficient to promote autonomous and profitable economic projects, expand the opportunity set faced by poor individuals and thereby alleviate poverty. Micro Finance Institutions fill a needed gap within the financial services by attending to poor and low-income earners.

Looking at the emerging theory of Micro Finance, recent developments in some developing countries have reinforced the contention that, Micro Finance structure is essential for the development of rural areas in consideration of the fact that areas of development in these countries have been traditionally urban-centered (Iheduru, 2002). The development of Micro Finance Institutions over the last few decades and their success has shown that Micro Finance is a major stimulus for combating poverty. Therefore Micro Finance as a strategy for economic development should be targeted at the poor given its multiplier effect on production and marginal propensity to consume. Access to credit by this group of people accelerates their income and equally increases their savings and consumption.

### 1.2.4 Credit Management

The credit management process begins with assessing the creditworthiness of the customer and his or her business viability. Nduta (2013) stressed that the first step in limiting credit risk involves screening clients to ensure that they have the willingness and ability to repay a loan. This is very important if the company chooses to extend some type of credit line or revolving credit to certain customers. Hence, some specific criteria are established to ensure that the customer or client is qualifying for that credit.

The credit management process involves certain factors or elements used to evaluate the customer for any form of financial services. This includes gathering data on the potential customer's current financial condition, including the current credit track record that discloses the character of a customer and collateral value. Hagos (2010). Sound credit management seeks to protect both the customers involved and the bank extending the credit from possible losses.

When the process of credit management functions effectively, everyone involves benefits from that and the chance of default is defeated. . Financial institutions should have a reasonable assurance that loans granted to a client will be paid back within the agreed terms. This can only be possible with the help of an effective credit policy. Credit policy can be seen as those decision variables that influence the amount of credit given out to customers. The formulation of the policy is the responsibility of the bank directors and management. The credit policy of any bank can be lenient or stringent depending on its approach; *Lenient Credit Policy*: Banks operating a Lenient credit policy tend to give credit facilities to customers very liberally that credit is granted even to those customers whose creditworthiness is not known or is doubtful. *Stringent Credit Policy*: These are banks that are very selective in extending credit or loans. They offer credit facilities to their customers who have proven creditworthiness. The banks with stringent policies follow tight credit standards and terms and as a result, minimize cost, risk, and chances of bad debts and also, the problem of liquidity.

### 1.2.5 Credit Management Variables

Sound credit management has the same certain key variables which the customers must meet before receiving the proposed credit arrangement. These are; client appraisal, credit terms, Credit Risk Control, and Collection Policy.

### 1.2.6 Collection Policy

The collection effort should therefore aim at accelerating collections from slow payers and reducing bad debt losses (Kariuki, 2010). Micro Finance Institutions can develop policies to ensure that credit management is carried out effectively since some of the customers are slow payers while others are non-payers.

**Figure 1: Relationship between Credit Management and Financial Performance**

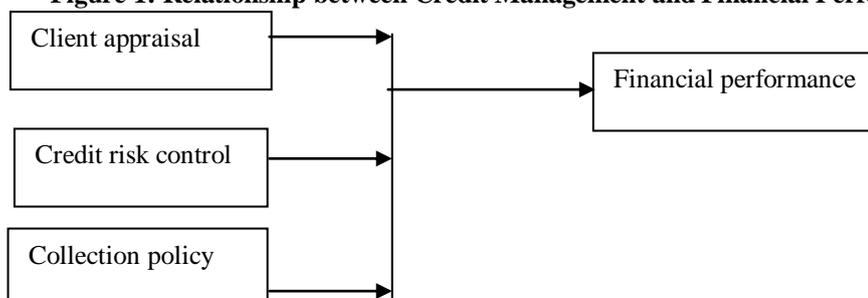

*Adapted from Haron, Justo, Nebat, and Sindani (2012)*

### 1.2.7 Variables Connectivity

Micro Finance Institution extends a loan to its client the interest rate to be charged on the loan and the credit period will be specified. Therefore, this will have an effect on their loan performance, since the time of repayment is clearly stated hence, reduces default. Client appraisal helps Micro Finance Institutions in knowing their potential borrower through the use of the 5Cs of credit. Once the 5Cs are properly applied the bank will know the creditworthiness and integrity of the borrower thereby reduce default. Credit risk, also known as default risk which is an investor's risk of loss arising from a borrower who does not make payment as promised. In such a situation it is called default. Credit risk control helps Micro Finance institutions to reduce the rate of nonpayment. The collection policy is needed because some customers are slow payers while others are non-payers. The collection policy aim at accelerating collections from slow and non-payers and reducing bad debt losses. Also, the policies encourage prompt payment and reduce the rate of default.

### 1.2.8 Portfolio Quality

A portfolio indicates total funds available for the MFIs to use as loans to their clients. Portfolio quality is very important to the financial success of MFIs. Drop-in portfolio quality could mean a drastic decline in customer satisfaction, which may lead to a low retention rate resulting in a higher cost to recruit new clients. The primary asset of MFI is its gross loan portfolio (Nelson, 2011) and in addition, the quality of that asset and therefore, the risk it poses for the institution can be quite difficult to measure. Portfolio quality is a critical area of performance analysis since the largest source of risk for any financial institution resides in its loan portfolio. For Micro Finance institutions, whose loans are not backed by collateral, the quality of the portfolio is absolutely crucial (AEMFI, 2013), poor asset quality will result in additional cost and lower income.

The portfolio quality indicator described the loan portfolio which is pivotal to financial institutions. The standard practice and regulation expectation is that MFI should calculate portfolio at risk (PAR) because if a payment is late the entire loan balance is at risk as the likelihood of recovering the balance of the loan is less even when payment is late. The indicator reflects the management's ability to recover loans disbursed but system lapses and high delinquency make financial sustainability difficult. Portfolio at Risk (PAR) is important because it indicates the potential for future loss based on the current performance of the loan portfolio the PAR ratio is the most widely accepted measure of loan performance in the microfinance institution. PAR>30 days is often used as the threshold beyond which loans are considered to be at higher risk. This ratio should be low and fairly stable and banks should monitor it daily, if possible when referring to PAR, the MFI should always specify the number of days. Calculating PAR.>1 day is an excellent management tool to monitor loan repayment and the risk of default it enables banks to address a problem before it gets out of control. Therefore, MFIs must try to maintain the quality of their portfolios. Portfolio quality is measured as a portfolio at risk over 30 days (Pr 30> days).

### 1.3 Research Objectives

The objective of this paper is to provide a systematic evaluation of the effect of collection policy on the Portfolio Quality of Micro-Finance Bank in Adamawa State.

### 1.4 Research Methodology and Data Analysis

We use regression analysis and descriptive statistics to analyze the data collected and to also test our proposed hypothesis. The proposed hypothesis is - Debt Collection Policy has no significant effect on the Portfolio Quality of Micro Finance banks. Furthermore, a multi-stage sampling method was used to select a sample of 21 respondents from a population of 51 credit officers.

This research study assesses the effect of collection policy on the portfolio quality of microfinance banks in Adamawa state for the years 2010 – 2016. These periods were selected because of the significant growth in small-scale businesses in Adamawa State which increased the demand for credit facilities. Moreover, it covered the eight (8) licensed Micro Finance Banks that were operating in Adamawa state for the period covered. We list the names of the banks as follows: Biyama Micro Finance Bank Limited, Hong LGA, Bonghe Micro Finance Bank Limited, Numan LGA, Fufore Micro-Finance Bank Limited, Fufore LGA, Girei microfinance bank Girei LGA, Gudusisa Micro Finance Bank Limited, Gombi LGA, Michika Micro-Finance Bank, Michika LGA, Standard Micro-Finance Bank Limited, Yola North LGA and Ummah Micro Finance Bank, Yola South LGA which covers the northern, central and southern senatorial district of Adamawa State.

**1.4.1 Data Presentation and Analysis:** A total number of fifty-one (51) questionnaires were distributed to the credit officers of seven selected microfinance banks in Adamawa State. They were filed and returned because of adequate time given to them. . However, prior to the test of each hypothesis, Pearson's correlation of both one and two-tailed significance was employed to explain the association between the regresses and the regressors as well as the association between the regressors themselves. Similarly, the Kolmogorov – Smirnov normality test was used to test whether the sample distribution is significantly different from a normal distribution. Finally, multiple regression was used to test the hypothesis.

**1.4.2 Data Presentation, Analysis, and Interpretation**: An empirical analysis and interpretation were done from the data collected through the designed questionnaires. They are presented serially in tables in accordance with each question as below.

**Table 1. Sex of Respondents**

|  | Frequency | Percent | Valid Percent | Cumulative Percent |
|---|---|---|---|---|
| **Male** | 37 | 72.5 | 72.5 | 72.5 |
| **Female** | 14 | 27.5 | 27.5 | 100.0 |

| | Frequency | Percent | Valid Percent | Cumulative Percent |
|---|---|---|---|---|
| **Male** | 37 | 72.5 | 72.5 | 72.5 |
| **Female** | 14 | 27.5 | 27.5 | 100.0 |
| **Total** | 51 | 100.0 | 100.0 | |

*Data Source: Survey Work 2020.*

Table 1 indicates the distribution of the respondents according to their sex. Respondents that are male constitute the majority with about 37 representing 72.5%. However female respondents accounted for the least with 14 denoting 27.5%. This shows that the majority of the respondents are male. This might not escape from the fact that most of our workforce are males. The fact that they are male more production is expected as men are less busy with household responsibility and maternity leaves and so on.

**Table 2. Age of Respondents**

| | Frequency | Percent | Valid Percent | Cumulative Percent |
|---|---|---|---|---|
| **18 - 25** | 11 | 21.6 | 21.6 | 21.6 |
| **26 - 33** | 21 | 41.2 | 41.2 | 62.8 |
| **34 - 41** | 10 | 19.6 | 19.6 | 82.4 |
| **42 - 49** | 5 | 9.8 | 9.8 | 92. |
| **50 +** | 4 | 7.8 | 7.8 | 100.0 |
| **Total** | 51 | 100.0 | 100.0 | |

*Data Source: Survey Work 2020.*

Table 2 shows the distribution of respondents according to age. Those who are in the age range of 26 – 33 years formed the majority of the respondents with 21 portraying 41.2%. This is seconded by those between the ages of 18 – 25 years with 11 representing 21.6%. Those within the age bracket of 34 – 41 years accounted for the third position with 10 representing 19.6%. However, those within the age bracket of 42 – 49 and 50 years above constituted the least with 5 and 4 respondents connoting 9.8% and 7.8%. This implies that most of the credit management administrators of the banks are in their active age. Therefore full productivity is expected in the respective microfinance banks.

**Table 3. Distribution of Respondents according to Name of Bank**

| Variable | Frequency | Percentage | Cumulative Percent |
|---|---|---|---|
| **Biyama** | 5 | 9.8 | 9.8 |
| **Bonghe** | 8 | 15.7 | 25.5 |
| **Furore** | 8 | 15.7 | 41.2 |
| **Girei** | 7 | 13.7 | 54.9 |
| **Gudusisa** | 8 | 15.7 | 70.6 |
| **Standard** | 8 | 15.7 | 84.3 |
| **Ummah** | 7 | 13.7 | 100.0 |
| **Total** | 51 | 100.0 | |

*Data Source: Survey Work 2020.*

Table 3 presents the names of banks of the respondents, with those coming from Bonghe, Furore, Gudusisa, and Standard Microfinance banks accounted for the highest with 8 denoting 15.7% each. Respondents coming from Girei and Ummah microfinance banks accounted for the second with 7 connoting 13.7% each. However, opinions coming from respondents from Biyama Microfinance recorded as the least with 5 revealing 23.8%. This means that the banks are proportionally or equally represented in respect of credit management.

**Table 4. The Existence of Microfinance Banks**

| Variable | Frequency | Percentage |
|---|---|---|
| Less than 5 years | 0 | 00.0 |
| Between 5 – 10 years | 13 | 25.5 |
| Between 10 – 15 years | 38 | 74.5 |
| Above 15 years | 0 | 00.0 |
| Total | 51 | 100.0 |

*Data Source: Survey Work 2020.*

Table 4 reveals how long the microfinance banks of the respondents were being in existence. Respondents, that confirmed that their bank is in existence between 10 – 15 years accounted for the highest with 38 people representing 74.5%. This is because microfinance banks started operation in Nigeria in the year 2005 and some of them were in existence before as community banks. However, 13 respondents denoting 25.5% attested that they were in the business between 5 – 10 years. This implies that almost all the banks were in existence for over 10 years therefore might have a well-structured credit management policy.

**Table 5. Adoption of Credit Management Practice**

| Variable | Frequency | Percentage |
|---|---|---|
| Yes | 51 | 100.0 |
| No | 0 | 00.0 |
| Total | 51 | 100.0 |

*Data Source: Survey Work 2020.*

Furthermore, Table 5 shows the adoption of credit management practices by microfinance banks. It indicates that all the banks adopt credit management practice with 51 respondents representing 100%. However, there is no response with regard to variable 'NO' which suggests the absence of credit management practice by the banks. This means that all the banks engage in credit management practice, perhaps the practice may vary from one bank to another.

**Table 6. Number of Clients by respective Microfinance Banks**

| Variable | Frequency | Percentage |
|---|---|---|
| Less than 100 | 4 | 7.8 |
| Between 100 - 200 | 2 | 3.9 |
| Between 200 - 300 | 1 | 2.0 |
| Above 300 | 44 | 86.3 |
| Total | 51 | 100.0 |

*Data Source: Survey Work 2020.*

Table 6 depicts the number of clients possesses by each microfinance bank in the state. Microfinance banks that possess more than 300 clients accounted for the highest with 44 respondents denoting 86.3%. Meanwhile, those with less than 100 clients have 4 representing 7.8%. However, those with 100 - 200 clients recorded as the second least with 2 revealing 3.9%. Those in between 200 - 300 recorded as the least with 1 each representing 2.0%. This means that majority of the banks have more than 300 clients, therefore credit management practice is essential in order to avoid default payment. Equally, it helps to maintain liquidity in the banks.

**Table 7: Success of Collection Policy of Microfinance Banks**

| Variable | Frequency | Percentage |
|---|---|---|
| Poor | 0 | 00.0 |
| Moderate | 12 | 23.5 |
| Strong | 39 | 76.5 |
| Total | 51 | 100.0 |

*Data Source: Survey Work 2020.*

Table 7 shows the score of the success of the collection policy of microfinance banks by the respondents. Those who scored the success of the collection policy as strong accounted for the highest with 39 representing 76.5%. Meanwhile, those who considered the success of the collection policy as moderate recorded as the second with 12 representing 23.5%. However, those who scored the success of the collection policy as poor accounted for nothing. This indicates that the collection policy of the banks was averagely excellent. A sound and well-articulated collection policy is an integral tool for credit management by banks.

**Table 8.  Portfolio Quality of the Microfinance Banks before Collection Policy**

| Variable | Frequency | Percentage |
|---|---|---|
| Strongly Agree | 0 | 00.0 |
| Agree | 0 | 00.0 |
| Neutral | 8 | 15.7 |
| Disagree | 19 | 37.3 |
| Strongly Disagree | 24 | 47.1 |
| Total | 51 | 100.0 |

*Data Source: Survey Work 2020.*

Table 8 displays the level of agreement of respondents on portfolio quality of the banks towards contribution to the overall goals of the banks before improvement in collection policy. Variable 'Strongly Disagree' accounted for the first position with 24 representing 41.7%. It was seconded by the variable 'Disagree' with 19 respondents revealing 37.3%. The third position goes to variable 'Neutral' with 8 respondents reflecting 15.7%. However, two variables 'Strongly Agree' and 'Agree' recorded zero each depicting nothing. This implies that the majority of the respondents are either strongly disagree or disagree with the portfolio quality of the banks.

**Table 9:  Available Collection Policies toward Effective Credit Management**

| Variable | Frequency | Percentage |
|---|---|---|
| Strongly Agree | 32 | 62.7 |
| Agree | 16 | 31.4 |
| Neutral | 3 | 5.9 |
| Disagree | 0 | 00.0 |
| Strongly Disagree | 0 | 00.0 |
| Total | 51 | 100.0 |

*Data Source: Survey Work 2020.*

Table 9 depicts the level of agreement of the respondents on available collection policies of their banks toward effective credit management. Those strongly agree that the level of available collection policies of their bank is effective on credit management, accounted for the highest with 32 representing 62.7%. This was followed by respondents who agree on the effectiveness of available collection policies on credit management with 16 representing 31.4%. However, only 3 respondents connoting 5.9% are neutral about the effectiveness of available collection policies on credit management. Variables 'Disagree' and 'Strongly Disagree' recorded zero scores each. This implies that the available collection policies of microfinance banks are strong. The goodness of available collection policies of the banks is a sign of effective credit management.

**Table 10. Developing Proper Collection Policies have been a Major Difficulty in Credit Management**

| Variable | Frequency | Percentage |
|---|---|---|
| Strongly Agree | 21 | 41.2 |
| Agree | 18 | 35.3 |
| Neutral | 5 | 9.8 |
| Disagree | 3 | 5.9 |
| Strongly Disagree | 4 | 7.8 |
| Total | 51 | 100.0 |

*Data Source: Survey Work 2020.*

Table 10 presents the level of agreement on developing proper collection policies have been a major difficulty in credit management. Variable 'Strongly Agree' marked as the highest with 21 respondents reflecting 41.2%. It was followed by the variable 'Agree' with 18 people denoting 35.5%. The third position goes to variables 'Neutral' with 5 respondents representing 9.8%. Two variables 'Strongly Disagree' and 'Disagree' accounted for the first and last position with 4 and 3 respondents representing 7.8% and 5.9%. This means that majority of the respondents agree that developing proper collection policies have been a major difficulty in credit management.

**Table 11. Proper adherence to Strict Collection Policies provides Chance for Loans Recovery**

| Variable | Frequency | Percentage |
|---|---|---|
| Strongly Agree | 33 | 64.7 |
| Agree | 15 | 29.4 |
| Neutral | 3 | 5.9 |
| Disagree | 0 | 00.0 |
| Strongly Disagree | 0 | 00.0 |
| Total | 51 | 100.0 |

*Data Source: Survey Work 2020.*

Table 11 describes the level of agreement of respondents on how proper adherence to strict collection policies provides a chance for loan recovery. Those that have strongly agree with the view accounted for the highest with 33 respondents representing 64.7%. It is seconded by respondents who agree with the notion with 15 denoting 29.4%. Only 3 respondents representing 5.9% are neutral about the idea that proper adherence to strict collection policies provides a chance for loan recovery. However, there is no score for respondents who either strongly disagree or disagree with the view. This means that the majority of the respondents strongly agree that proper adherence to strict collection policies provides a chance for loan recovery.

**Table 12. Staff Motivation is Effective in Improving Recovery of Delinquent Loans**

| Variable | Frequency | Percentage |
|---|---|---|
| Strongly Agree | 28 | 54.9 |
| Agree | 17 | 33.3 |
| Neutral | 6 | 11.8 |
| Disagree | 0 | 00.0 |
| Strongly Disagree | 0 | 00.0 |
| Total | 51 | 100.0 |

*Data Source: Survey Work 2020.*

Table 12 shows the level of agreement by respondents on how staff motivation is effective in improving the recovery of delinquent loans. Variable 'Strongly Agree' accounted for the highest with 28 respondents representing 54.9%. This is closely followed by the variable 'Agree' with 17 representing 33.3%. Variable 'Neutral' recorded the third score with 6 connoting 11.8%. However variables 'Disagree' and 'Strongly Disagree' recorded nothing, indicating that majority of the respondents have strongly agreed that staff motivation is effective in improving recovery of delinquent loans.

**Table 13. Frequent Review of Collection Policies as a Tool for Improving Credit Management**

| Variable | Frequency | Percentage |
|---|---|---|
| Strongly Agree | 30 | 58.8 |
| Agree | 17 | 33.3 |
| Neutral | 4 | 7.8 |
| Disagree | 0 | 00.0 |
| Strongly Disagree | 0 | 00.0 |
| Total | 51 | 100.0 |

*Data Source: Survey Work 2020.*

Table 13 presents the level of agreement of respondents on how frequent reviews of collection policies improve credit management. Those that strongly agree accounted for the highest with 30 representing 58.8%. It is closely followed by those who agree with the view recording 17 denoting 33.3%. Only 4 respondents representing 7.8% who remain neutral about frequent reviews of collection policies on improving credit management. However, those who have strongly disagree or disagree accounted for nothing, depicting that frequent reviews of collection policies help in improving credit management.

**Table 14. Stringent policy is more effective in Debt Recovery than a Lenient Policy**

| Variable | Frequency | Percentage |
|---|---|---|
| Strongly Agree | 27 | 52.9 |
| Agree | 12 | 23.5 |
| Neutral | 7 | 13.7 |
| Disagree | 5 | 9.8 |
| Strongly Disagree | 0 | 00.0 |
| Total | 51 | 100.0 |

*Data Source: Survey Work 2020.*

Table 14 explains the effectiveness of the stringent policy on debt recovery when compared with lenient policy. Respondents who strongly agree that stringent policy is more effective in debt recovery than a lenient policy accounted for the highest with 27 representing 52.9%. This is seconded by respondents who agree with the view recording 12 constituting 23.5%. The third position goes to respondents who considered it neutral with 7 connoting 13.7%. However, only 5 respondents representing 9.8% disagree with the notion, while variable strongly disagrees accounted for nothing, meaning that stringent policy is more effective in debt recovery than a lenient policy.

Test of Hypothesis
Hypothesis: Debt Collection Policy has no significant effect on Portfolio Quality of Microfinance Banks.

**Table 15. Model Summary 1**

| Model | R | R Square | Adjusted R Square | Std. Error of the Est. |
|---|---|---|---|---|
| 1 | 0.781 | 0.610 | 0.556 | 0.52090 |

*Data Source: SPSS Version 17 Computation*

Table 15 reveals that 'R' reflects the relationship between the independent variable (available collection policies toward effective credit management, developing proper collection policies as a major difficulty in credit management, proper adherence to collection policies provides a chance for loan recovery, staff motivation is effective in improving recovery of delinquent loans, frequent review on collection policies help to improve credit management and stringent policy is more effective in debt recovery than a lenient policy) and dependent variable (portfolio quality of microfinance banks). The rule establishes that, the closer the figure to 1 the stronger the relationship and vice versa. Therefore with reference to this model, the relationship between the independent and dependent variable is good and strong with 0.781. This implies that the model has a robust and strong goodness fit; depicting that available collection policies toward effective credit management, developing proper collection policies as a major difficulty in credit management, proper adherence to collection policies provides a chance for loan recovery, staff motivation is effective in improving recovery of delinquent loans, frequent review on collection policies help to improve credit management and stringent policy is more effective in debt recovery than a lenient policy has a positive effect on portfolio quality of microfinance banks. Furthermore, the model presents R2 of 0.610, implying that about 61% increase in the portfolio quality of microfinance banks is accounted for by the variables in the model while the remaining 39% is accounted for by other factors not captured by the model. The robustness and goodness of fit of the model are further confirmed by adjusted R2 of 0.556, which means that about 56% of the variation in the dependent variable is accounted for by the regresses.

**Table 16. Model Summary 2**

| Model | Sum of Squares | DF | Mean Square | F | Sig. |
|---|---|---|---|---|---|
| Regression | 18.264 | 6 | 3.108 | 11.455 | 0.000 |
| Residual | 11.339 | 44 | 0.271 | | |
| Total | 30.588 | 50 | | | |

*Data Source: SPSS Version 17 Computation*

Table 16 utilizes F – Statistics to tests for the overall significance of the hypothesis and regressor of the study. The regressor is significant at both 1% and 5% levels of significance. The rule says if the calculated F value is greater than the tabulated F value rejects the null hypothesis. Therefore, with regard to this model the F calculated is 11.455 while F tabulated or significant value is 0.000. Since F calculated is greater than F tabulated, so we reject the null hypothesis of no significance and infer that debt collection policy has a significant effect on the portfolio quality of microfinance banks.

**Table 17. Model Summary 3**

| Model | Coefficient | T – Statistics | Sig. Value |
|---|---|---|---|
| 1 (Constant) | 5.264 | 18.414 | 0.000 |
| Effectiveness of available collection policies on credit Mgt. | -0.565 | -4.373 | 0.000 |
| Difficulty in developing proper collection policies in Credit Mgt. | 0.072 | 0.551 | 0.591 |
| Strict collection policies provide chances for loan recovery | 0.331 | 0.887 | 0.390 |
| Staff motivation is effective in the recovery of a delinquent loan | 0.084 | 0.366 | 0.720 |
| Frequent reviews on collection policies improve the level of credit Mgt. | 0.623 | 2.074 | 0.057 |
| The stringent policy is more effective in debt recovery than a lenient policy | -0.059 | -0.323 | 0.751 |

*Data Source: SPSS Version 17 Computation*

Table 17 depicts the effects of the independent variable (available collection policies toward effective credit management, developing proper collection policies as a major difficulty in credit management, proper adherence to strict collection policies provides a chance for loan recovery, staff motivation is effective in improving recovery of delinquent loans, frequent review on collection policies help to improve credit management and stringent policy is more effective in debt recovery than a lenient policy) and dependent variable (portfolio quality of microfinance banks). But the magnitude of the effects is of varying degree. Frequent review on collection policies has a strong positive effect given a coefficient of 0.623. This implies that frequent review on collection policies helps to improve the level of credit management, and thereby bringing about a change in portfolio quality of microfinance banks of 62%. Similarly, proper adherence to strict collection policies has an average positive effect given a coefficient of 33%. This means that strict collection policies provide chances for loan recovery by about 33%. Moreover, staff motivation and developing proper collection policies in credit management depict positive coefficients of 0.084 and 0.072. This indicates that staff motivation is effective in improving recovery of delinquent loans by about 8%, and developing proper collection policies have been a major difficulty in credit management with about 7%. However, based on prior knowledge of the above table, available collection policies and stringent debt recovery showed negative effects with regards to credit management. Although the negative effects exerted by such variables were revealed to be negligible, marginal and low, given coefficients of -0.059 and -0.565. The multiple regression results have also revealed that frequent review on collection policies is the most statistically significant of the regressors with a T – statistics of 2.074 at a 1% level of significance. A strict collection policy by the banks is also statistical significance with a T – statistics of 0.887 at a 1% level of significance. Equally difficult in developing proper collection policies is also statistical significance with a T – statistics of 0.551 at a 1% level of significance. However stringent collection policies and available collection policies were found to be statistically insignificant at a 1% level of significance given a T – statistics of -0.323 and -4.373. But the general outcome of the multiple regression models (results) showed a positive effect between the independent and dependent variables.

## II.    CONCLUSION

Based on the findings of this study, it can be concluded that credit management plays a positive significant role in improving the financial performance of microfinance banks. This is attributed to the fact that sounds and grounded credit management (client appraisal) allowed the bank to be efficient and have availability of liquidity. The performance or strength of any bank is judge by its liquidity. This is because the efficiency of any bank tied up with the availability of cash. Meanwhile, the profitability of microfinance banks has improved. This is a result of proper credit risk control measures adopted by the banks. Credit risk control measures allowed the banks to give loans to only clients who worth it, thereby helping them to generate more profit. Lastly, with regard to the portfolio quality of microfinance banks, the study concluded that credit management plays a significant role in improving it. However frequent review of collection policy was proved to be very effective in credit management. Moreover, this study recommends that microfinance banks should keep on reviewing their debt collection policy as frequent review of debt collection policy in credit management is very effective in improving the portfolio quality of the banks. Banks should keep on monitoring and evaluating the proper functioning of the credit management section as the unit proves to be very significant in improving the financial performance of the banks.